# Dispersive charge density wave excitations and temperature dependent commensuration in $Bi_2Sr_2CaCu_2O_{8+\delta}$

L. Chaix[1], G. Ghiringhelli[2,3], Y. Y. Peng[2], M. Hashimoto[4], B. Moritz[1], K. Kummer[5], N. B. Brookes[5], Y. He[6], S. Chen[6], S. Ishida[7], Y. Yoshida[7], H. Eisaki[7], M. Salluzzo[8], L. Braicovich[2,3], Z.-X. Shen[1,6,*], T. P. Devereaux[1,6,*], W.-S. Lee[1,*]

[1]Stanford Institute for Materials and Energy Sciences, SLAC National Accelerator Laboratory and Stanford University, 2575 Sand Hill Road, Menlo Park, California 94025, USA.

[2]Dipartimento di Fisica, Politecnico di Milano, Piazza Leonardo da Vinci 32, I-20133 Milano, Italy.

[3]CNR-SPIN, CNISM, Politecnico di Milano, Piazza Leonardo da Vinci 32, I-20133 Milano, Italy.

[4]Stanford Synchrotron Radiation Lightsource, SLAC National Accelerator Laboratory, 2575, Sand Hill Road, Menlo Park, California 94025, USA.

[5]European Synchrotron Radiation Facility (ESRF), BP 220, F-38043 Grenoble Cedex, France.

[6]Geballe Laboratory for Advanced Materials, Stanford University, Stanford, California 94305, USA.

[7]National Institute of Advanced Industrial Science and Technology (AIST), Japan.

[8]CNR-SPIN, Complesso Monte Sant'angelo, Via Cinthia, I-80126 Napoli, Italy.

*Correspondence to: leews@stanford.edu, zxshen@stanford.edu, tpd@stanford.edu

**Experimental evidence on high-*Tc* cuprates reveals ubiquitous charge density wave (CDW) modulations[1-10], which coexist with superconductivity. Although the CDW had been predicted by theory[11-13], important questions remain about the extent to which the CDW influences lattice and charge degrees of freedom and its characteristics as functions of doping and temperature. These questions are intimately connected to the origin of the CDW and its relation to the mysterious cuprate pseudogap[10,14]. Here, we use ultrahigh resolution resonant inelastic x-ray scattering (RIXS) to reveal new CDW character in underdoped $Bi_2Sr_2CaCu_2O_{8+\delta}$. At low temperature, we observe dispersive excitations from an incommensurate CDW that induces anomalously enhanced phonon intensity, unseen using other techniques. Near the pseudogap temperature *T\**, the CDW persists, but the associated excitations significantly weaken and the CDW wavevector shifts, becoming nearly commensurate with a periodicity of four lattice constants. The dispersive CDW excitations, phonon anomaly, and temperature dependent commensuration provide a comprehensive momentum space picture of complex CDW behavior and point to a closer relationship with the pseudogap state.**

With sufficient energy resolution, RIXS can be an ideal probe for revealing the CDW excitations in cuprates. By tuning the incident photon energy to the Cu $L_3$-edge (Fig. 1a), the resonant absorption and emission processes can leave the system in excited final states, which couple to a variety of excitations arising from orbital, spin, charge, and lattice degrees of freedom[15]. Thus, information of these elementary excitations in energy and momentum space can be deduced from analyzing the RIXS spectra as functions of the energy loss and the momentum transfer of the photons (Fig. 1a). This is highlighted by the pivotal role that RIXS has recently played in revealing orbital and magnetic excitations in cuprates[16-20]. In addition, RIXS provided the first x-ray scattering evidence for an incommensurate CDW in the $(Y,Nd)Ba_2Cu_3O_{6+\delta}$ [4], owing to energy resolution that separated the quasi-elastic CDW signal (bright spot in Fig. 1b, limited by the instrumental resolution ~ 130 meV) from other intense higher-energy excitations. Notably this quasi-elastic signal is asymmetric with respect to zero energy loss (Fig. 1c), which indicates the possible existence of additional low energy excitations near the CDW wavevector ($Q_{CDW}$).

In this work, we exploit the newly commissioned ultrahigh resolution RIXS instrument at the European Synchrotron Radiation Facility to reveal these low energy excitations near the CDW. We choose the double-layer cuprate $Bi_2Sr_2CaCu_2O_{8+\delta}$ (Bi2212), whose electronic structure has been extensively studied by surface-sensitive spectroscopy, such as scanning tunneling microscope[21] and angle-resolved photoemission[22] and in which a short-range CDW order was recently reported[7,8]. With improved energy resolution up to 40 meV, we see additional features in the previous quasi-elastic region (Fig. 1c).

Figure 2a presents an energy-momentum RIXS intensity map of our high resolution data. Two excitation branches are clearly observed with a momentum dependent intensity distribution, also evident in the energy-loss spectra at representative momentum (Fig. 2b). The first branch centered at zero energy should contain the scattering signal due to an underlying CDW. Indeed, as plotted in Fig. 2c, the momentum-distribution of the RIXS intensity averaged over a small energy window exhibits a symmetric peak at a finite momentum with a peak-to-background ratio of ~ 2, unambiguously confirming the existence of the CDW in Bi2212. The peak position is located at $Q_{CDW} \sim 0.3$ reciprocal lattice units (r.l.u.), consistent with previous STM studies[8] and with a full width at half maximum of approximately 0.085 r.l.u., corresponding to a short correlation length of approximately 15 Å. Interestingly, despite of the short correlation length, the CDW strength, estimated by the integrated intensity of the CDW peak in the quasi-elastic region, is not weaker than that in $YBa_2Cu_3O_{6+x}$ (See Supplementary Text).

The second branch of excitations possesses an energy scale of approximately 60 meV, whose energy-momentum-dispersion can be reliably extracted (Fig. 2d and Supplementary Fig. 2). The extracted dispersion agrees well with that of the Cu-O bond stretching phonon measured by non-resonant inelastic x-ray scattering[23] (Supplementary Text). Owing to the high momentum resolution of our data, Fig. 2e shows mode softening by approximately 25% at $Q_{CDW}$, with a corresponding broadening of the fitted peak width, indicating that the short-ranged ordered CDW unambiguously affects the lattice. This observation is reminiscent of the bond-stretching phonon anomaly reported in the striped ordered cuprates, where CDW correlation lengths are notably longer[24].

Interestingly, the phonon intensity varies non-monotonically with momentum, having a maximum near $Q_A \sim 0.37$ r.l.u. (Fig. 3a). Examining the momentum distribution curves (Fig. 3b), the peak at zero energy disperses and broadens with increasing energy loss, smoothly connecting to the phonon at $Q_A$. To better visualize this connection, we remove the elastic contribution from the energy-loss spectra (Fig. 3c) to reveal a funnel-shaped RIXS intensity emanating from $Q_{CDW}$, suggesting that these excitations are associated with the CDW. It appears that these CDW excitations disperse to high energy and intersect with the phonon at $Q_A$, causing the phonon intensity anomaly.

It is important to note that the RIXS phonon cross-section directly reflects the momentum-dependence of the electron-phonon coupling strength[25-27] (Supplementary Text), in stark contrast with the phonon self-energy measured by other scattering techniques. As a consequence, RIXS is also directly sensitive to the interference effect between phonons and underlying charge excitations, *i.e.* the Fano effect, which can manifest as an intensity anomaly. Figure 3f illustrates this point, showing a calculated RIXS intensity map for a 1-D metallic system with coupling between electrons and a bond stretching phonon mode. Perfect Fermi surface nesting at $2k_F$ (*i.e.* twice the Fermi momentum) creates a particle-hole continuum, whose presence softens the phonon dispersion and creates an intensity anomaly at the intersection between the continuum and the phonon (Fig. 3d–f). This result qualitatively agrees with our observations in Bi2212, lending further support to the notion that some form of charge excitations associated with the CDW exists in underdoped Bi2212 and causes the observed phonon anomaly. By connecting $Q_{CDW}$ and $Q_A$, we deduce a characteristic velocity of these CDW excitations: $V_{CDW} \sim 0.6 \pm 0.2$ eV•Å. We find that $V_{CDW}$ is neither equivalent to the Fermi velocity seen in ARPES nor the quasi-particle-interference dispersion seen in STM[21] (Supplementary Text), indicating that the weak coupling fermiology may be irrelevant for describing the observed dispersion.

Temperature dependent measurements were conducted on another Bi2212 crystal with a similar $T_c$ (40 K) with results shown in Fig. 4. We first confirm that the low temperature measurement at 20 K reproduces all the observations shown in Fig. 2 (Supplementary Fig. 3). Upon increasing the temperature to 240 K (near the pseudogap temperature[22]), we find that the CDW signal in the quasi-elastic region persists (Fig. 4a). Surprisingly, $Q_{CDW}$ changes from incommensurate 0.3 r.l.u. at 20 K to nearly commensurate ~ 0.26 r.l.u. at 240 K (Fig. 4b and c), which corresponds to a real-space periodicity of approximately four lattice constants (~$4a_o$), similar to the stripe ordered cuprates[1,3]. We also observed that the intensity anomaly at $Q_A$ diminishes and broadens, as shown by the momentum-distribution of the phonon intensity in Fig. 4d. This indicates that the CDW excitations become poorly defined in the energy-momentum space along with the weakening of the CDW.

Our results have several important implications. They provide the first direct evidence for the existence of dispersive CDW excitations in the charge-charge correlation function with a characteristic velocity $V_{CDW}$. Furthermore, these excitations persist to high energy, at least up to ~ 60 meV, causing the phonon intensity anomaly, indicating that the CDW influences the charge and lattice degrees of freedom, despite its short correlation length. Whether the observed CDW excitations are collective due to a symmetry breaking[28] or rather a reflection of the particle-hole continuum remains an open question for future investigation. Additionally, the surprising change of $Q_{CDW}$ from incommensurate at low temperature toward commensurate at high temperature is reminiscent of the temperature dependence observed in striped phase nickelates where the charge order is strongly coupled to a spin order, which originates from strong real-space interactions[29]. As a function of doping, a similar incommensurate-to-nearly-commensurate transition is found across $p$ ~ 10% from low temperature STM studies on Bi2212 [8]. Our temperature dependent observations at a lower doping concentration ($p$ ~

8–9%) strongly support the existence of a universal, commensurate $4a_o$ CDW throughout the phase diagram of Bi2212. These results also are consistent with the idea that the incommensurate $Q_{CDW}$ forms from a discommensuration of the period $4a_o$ modulations[30]. Finally, the CDW survives up to at least 240 K ~ $T^*$, which taken together with the similarities to striped nickelates indicates a close relationship between the CDW, spins, and the pseudogap.

**Method:**

**Sample growth, preparation and characterization**

The high quality UD-Bi2212 ($Bi_{2.2}Sr_{1.8}Ca_{0.8}Dy_{0.2}Cu_2O_{8+\delta}$) single-crystals (Supplementary Fig.1a, sample 1 and 2) were grown by floating-zone (FZ) methods and annealed in Nitrogen atmosphere. The samples were characterized and roughly aligned using Laue diffraction before our RIXS measurements (Supplementary Fig. 1a). The samples were cleaved in air right before loading into the high vacuum measurement chamber. The quality and the doping of the UD-Bi2212 samples ($p \sim 8$–9%) were verified by measuring the $T_C$ of the samples. The $T_C$ are 45 K and 40 K for sample 1 and 2, respectively (Supplementary Fig. 1b). The x-ray absorption curves for sample 1 and 2 are shown in Supplementary Fig. 1c, showing no apparent contamination. Data shown in Fig. 2 and 3 were taken on Sample 1. Data in Fig. 4 were taken on sample 2.

**Low-resolution RIXS measurements on UD-NBCO**

The low energy resolution RIXS data on the UD-NBCO ($Nd_{1.2}Ba_{1.8}Cu_3O_{6+\delta}$) sample were taken at the ADRESS beamline of the Swiss Light Source at the Paul Scherrer Institut (PSI, Switzerland) using the SAXES spectrometer[31,32]. The RIXS spectra were measured at $T = 60$ K, just below $Tc = 65$ K to maximize the CDW peak intensity and using linear vertical (σ) polarized incident photons. The total energy resolution was ~130 meV, the scattering angle was 130°. The RIXS map is obtained by combining spectra measured every 2° in the rotation of the incidence angle on the sample surface. The NBCO films, 100 nm think, were deposited by high oxygen pressure diode sputtering on $SrTiO_3$ (100) single crystals. The lattice parameters measured by x-ray diffraction are $a = b = 3.84$ Å and $c = 11.7$ Å.

**Ultrahigh resolution RIXS measurements on UD-Bi2212**

The RIXS measurements on the UD-Bi2212 sample 1 and 2 were performed using the ERIXS spectrometer at the ID32 beamline of the European Synchrotron Radiation Facility (ESRF, France). The RIXS spectra were taken with the photon energy of the incident x-rays tuned to the maximum of the absorption curve near the Cu $L_3$-edge (Supplementary Fig. 1c). The scattering geometry is sketched in Supplementary Fig. 1d. The data were collected at $T = 20$ K for the sample 1 and $T = 20$ K / 240 K for the sample 2 with a linear vertical polarization (σ-polarization) of the incident beam. The energy resolution was approximately $\Delta E \sim 40$ meV (sample 1) and $\Delta E \sim 45$ meV (sample 2). The scattering angle of the endstation was set at $2\theta = 149.5°$. Since the electronic state in Bi2212 is quasi-two-directional, *i.e.* rather independent along the *c*-axis, the data shown in this report are plotted as a function of projected momentum transfer $Q_{//}$, *i.e.* projection of $Q = k_f - k_i$ on the $CuO_2$ plane. Different in-plane momentum transfers, $Q_{//}$ (projection of the scattering vector $Q$ along [100]), were obtained by rotating the samples around the vertical *b*-axis. Note that the scattering vector $Q$ is denoted using the pseudotetragonal unit cell with $a = b = 3.82$ Å and $c = 30.84$ Å, where the *c*-axis is normal to the sample surface. In our convention, positive $Q_{//}$ corresponds to grazing-emission geometry and negative $Q_{//}$ corresponds to grazing-incidence geometry.

**Data analysis and fitting**

The ultrahigh resolution data (UD-Bi2212) were normalized to $I_0$ and corrected for self-absorption effects using the formalism described in Ref. 33. The zero energy positions were determined by comparing the spectrum recorded on small amount of silver paint (at each $Q_{//}$) near the sample surface. For more accuracy on the zero energy alignment, the data were fitted using a model involving a Gaussian (elastic peak), a Lorentzian (phonon peak) and an anti-symmetrized Lorentzian (paramagnon) convoluted with the energy resolution (Gaussian convolution). The results of these fits are presented on Supplementary Fig. 2 for the sample 1 ($T$ = 20 K) and Fig. 4b for the sample 2 ($T$ = 20 K and $T$ = 240 K).

From this analysis, we extracted the dispersion (peak position) and measured FWHM (full width at half maximum) of the phonon peak at $T$ = 20 K in the sample 1 (Fig. 2e) and at $T$ = 20 K and $T$ = 240 K in the sample 2 (Supplementary Fig. 3c and d). The dispersions and FWHM found at $T$ = 20 K for both sample 1 and sample 2 are similar, with the presence of a softening and a broadening around the CDW wavevector ($Q_{CDW}$ ~ 0.3 r.l.u.). The reproducibility of the low temperature data in two independent measurements and samples confirms our observations of the CDW, its excitations, and the intensity anomaly of the phonon (Supplementary Fig. 3a). At a high temperature ($T$ = 240 K), the phonon softening is shifted to a lower wavevector. In addition, we note that the phonon intensity extracted from the fitting procedure agrees with the averaged intensity around the phonon energy (Supplementary Fig. 4b).

**Theory**

The calculations depicted in Fig. 3d–f were carried out using a recent framework for phonon contributions to RIXS given in Ref. 27. Details are given in that reference. The theory developed in Ref. 27 considered lowest order diagrams for RIXS in the weak electron-phonon limit, giving the first contribution to RIXS coming from the emission of one phonon. The strength of the signal is directly tied to the momentum dependent electron-phonon coupling. In the limit of small coupling, the intensity smoothly varies in momentum space, with a position that closely follows the bare phonon dispersion. However, if the phonon dispersion crosses the electron-hole continuum, a Fano effect occurs leading to a strong interference between the dispersions of the two excitations, as shown in Fig. 3d–f.

Since we have used a simple 1D metallic model having a Peierls instability at $2k_F$, the particle-hole excitations disperse in energy and momentum in a cone from zero energy, with a dispersion set by the Fermi velocity. When the cone of charge excitations crosses the phonon line, an interference occurs that exploits the momentum dependent electron-phonon coupling $g(q)$. For the case of the bond-stretching phonon, $g(q)$ determined from the nature of deformation coupling between Cu and O in-plane displacements is largest for large momentum transfers. Therefore the intensity is stronger on the higher momentum side of the intersection of the phonon line with the charge excitations. This effect well reproduces the experimental observations.

**Data availability:**

The data that support the plots within this paper and other findings of this study are available from the corresponding authors upon reasonable request.

**Acknowledgments:**

We thank S. A. Kivelson for discussions. This work is supported by the U.S. Department of Energy (DOE), Office of Science, Basic Energy Sciences, Materials Sciences and Engineering Division, under contract DE-AC02-76SF00515. L.C. acknowledges the support from Department of Energy, SLAC Laboratory Directed Research and Development funder contract under DE-AC02-76SF00515. The data in Fig. 1b were carried out partly at the Advanced Resonant Spectroscopies (ADRESS) beam line of the Swiss Light Source, using the Super Advanced X-ray Emission Spectrometer (SAXES) instrument jointly built by Paul Scherrer Institut (Villigen, Switzerland), Politecnico di Milano (Italy), and École Polytechnique Fédérale de Lausanne (Switzerland); All other RIXS data were taken at the ID32 of the ESRF (Grenoble, France) using the ERIXS spectrometer designed jointly by the ESRF and Politecnico di Milano. ARPES data were taken at Stanford Synchrotron Radiation Lightsource is operated by the U.S. Department of Energy, Office of Science, Office of Basic Energy Sciences.

**Author Contributions:**

W.S.L., G.G., L.B., T.P.D. and Z.X.S. conceived the experiment. L.C., W.S.L., G.G., Y.Y.P., M.H., L.B. K.K., and N.B.B. conducted the experiment at ESRF. L.C., W.S.L., G.G., Y.Y.P., L.B. and M.H. analyzed the data. T.P.D. and B.M. performed the theoretical calculations. Y.H., S.C., S.I., Y.Y., H.E., and M.S. synthesized and prepared samples for the experiments. L.C. and W.S.L. wrote the manuscript with input from all the authors.

**Competing financial interests**

The authors declare no competing financial interests.

**Corresponding authors**

Correspondence to W. S. Lee (leews@stanford.edu), T. P. Devereaux (tpd@stanford.edu), or Z. X. Shen (zxshen@stanford.edu).

**Figure Caption:**

**Figure 1| RIXS process and a hint of lower energy excitations near $Q_{CDW}$. a**, Schematics of the coherent two-step RIXS process. The photon energy of the incident x-ray ($h\nu_i$) is tuned to the Cu $L_3$-edge. Upon absorption, a core electron in the Cu $2p$ orbital made a transition to the unoccupied $3d$ valence band near the Fermi energy $E_F$, bringing the system to an intermediate state $|m\rangle$. Then, the emission process occurs that fills the $2p$ core hole by one of the valence electrons, emits an photon ($h\nu_f$), and leaves the system in an excited final state $|f\rangle$, which couple to a variety of elementary excitations. The

energy-momentum information of these elementary excitations can be deduced by tracking the peak and spectral weight in RIXS spectra as a function of the momentum transfer $Q$ and the energy loss of the scattering photons. $k_i$ and $k_f$ represent the momentum of the incident and scattered photons, respectively. Since the electronic state in Bi2212 is quasi-two-directional, *i.e.* rather independent along the *c*-axis, all data are plotted as a function of projected momentum transfer $Q_{//}$ along [100] direction, *i.e.* along the Cu-O bond direction. **b**, RIXS intensity map of the underdoped $Nd_{1.2}Ba_{1.8}Cu_3O_{6+\delta}$ (UD-NBCO) compound around the CDW position, taken with a resolution of 130 meV. The red dashed line indicates the RIXS spectrum shown in **c**. **c**, Comparison of two RIXS spectra recorded with the low (UD-NBCO, $\Delta E \sim 130$ meV) and high energy resolution (UD-Bi2212, $\Delta E \sim 40$ meV) at $Q_{//} = 0.31$ r.l.u..

**Figure 2| CDW and phonons at 20K. a**, RIXS intensity map of the high energy resolution data ($\Delta E \sim 40$ meV) as a function of energy loss and $Q_{//}$. The white circles represent the fitted phonon dispersion. **b**, Energy loss spectra at selected momentum ranging from $Q_{//} = 0.235$ r.l.u. to $Q_{//} = 0.425$ r.l.u.. The fits (solid lines) are superimposed to the raw data (black circles). **c**, Averaged intensity in the quasi-elastic region, defined as the region between the two white dashed lines in **a**. The solid line is a Lorentzian fit to the data with a background consisting of a constant plus a Lorentzian to account for the tail of specular reflection peak at $Q_{//} = 0$. **d**, Demonstration of the quality of the fit of a RIXS spectrum at a representative momentum. The fitting process is described in the Method section. The phonon peak is highlighted in pink. **e**, Position and FWHM (full width at half maximum) of the measured phonon peak extracted from the fits. The red dashed lines on **a** and **e** indicate the $Q_{CDW}$. Error bars in **a** and **e** are estimated by the

uncertainty in determining the zero energy loss. Those in **c** are determined by the noise level of the spectra.

**Figure 3| CDW excitations in Bi2212 and calculated RIXS intensity in a 1D model. a**, RIXS phonon intensity extracted from the averaged intensity over the energy range between the two black dashed lines in **c**. The black arrow highlights the intensity anomaly. Error bars are estimated by the noise level of the spectra. **b,** Momentum distribution curves **(**MDCs**)** at fixed energy of the raw data (Fig. 2a) from $E = 0$ eV to $E = 0.1$ eV. Both the raw data (markers with dashed lines) and the smoothed curves (solid lines) are superimposed. The MDCs corresponding to the CDW and phonon energy regions are highlighted in blue. The red ticks serve as guides-to-the-eye for the connection between CDW and the phonon intensity anomaly. $Q_{CDW}$ represents the CDW wavevector. **c**, Elastic peak-subtracted intensity map. Calculations of RIXS phonon intensity for a 1D metallic system is shown in **d**, **e**, and **f**, with the same plotting format and notations used in panels **a**–**c**. $2k_F$ represents the perfect nesting wavevector of the 1D system. All the experimental (calculated) data are plotted as function of $Q_{//} - Q_{CDW}$ ($Q_{//} - 2k_F$).

**Figure 4| Temperature dependence of the CDW and the phonon anomaly. a**, RIXS intensity map taken at $T = 240$ K. The white dashed lines define the quasi-elastic region (centered at zero energy loss) and the phonon energy region (centered at 60 meV) used for calculating the average intensity profile shown in **c** and **d**. **b**, Raw energy loss spectra (markers) and the corresponding fits (solid lines) of RIXS data taken at $T = 240$ K (top) and $T = 20$ K (bottom). The elastic peak fits are highlighted with filled areas. The RIXS spectra where the elastic peak intensity is maximal (*i.e.* the CDW position) are indicated

by red and blue boxes for $T$ = 240 K and $T$ = 20 K, respectively. **c**, Averaged intensity of the quasi-elastic region (indicated in **a**) at 240 K (red) and 20 K (blue). The dashed lines indicate the position of the CDW. The RIXS intensity map taken at $T$ = 20 K is presented in the Supplementary Fig. 3. **d**, Averaged intensity of the phonon energy regions, as defined in **a**, at 240 K and 20 K. The black arrow highlights the intensity anomaly at 20 K. Error bars in **c** and **d** are estimated by the noise level of the spectra.

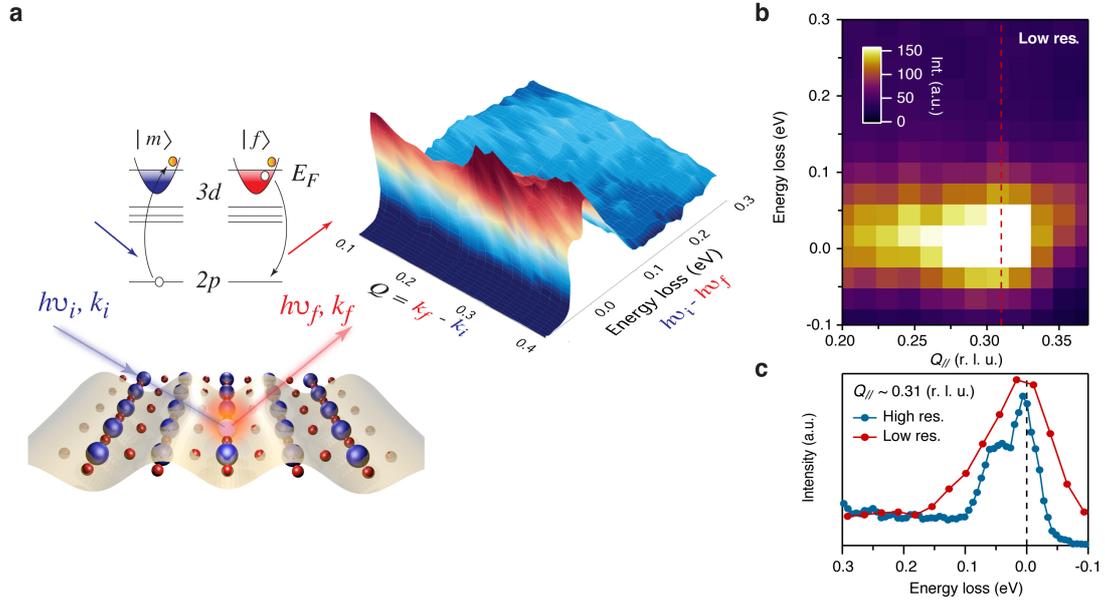

**Figure 1| RIXS process and a hint of lower energy excitations near $Q_{CDW}$. a**, Schematics of the coherent two-step RIXS process. The photon energy of the incident x-ray ($h\upsilon_i$) is tuned to the Cu $L_3$-edge. Upon absorption, a core electron in the Cu $2p$ orbital made a transition to the unoccupied $3d$ valence band near the Fermi energy $E_F$, bringing the system to an intermediate state $|m\rangle$. Then, the emission process occurs that fills the $2p$ core hole by one of the valence electrons, emits an photon ($h\upsilon_f$), and leaves the system in an excited final state $|f\rangle$, which couple to a variety of elementary excitations. The energy-momentum information of these elementary excitations can be deduced by tracking the peak and spectral weight in RIXS spectra as a function of the momentum transfer $Q$ and the energy loss of the scattering photons. $k_i$ and $k_f$ represent the momentum of the incident and scattered photons, respectively. Since the electronic state in Bi2212 is quasi-two-directional, *i.e.* rather independent along the *c*-axis, all data are plotted as a function of projected momentum transfer $Q_{//}$ along [100] direction, *i.e.* along the Cu-O bond direction. **b**, RIXS intensity map of the underdoped $Nd_{1.2}Ba_{1.8}Cu_3O_{6+\delta}$ (UD-NBCO) compound around the CDW position, taken with a resolution of 130 meV. The red dashed line indicates the RIXS spectrum shown in **c**. **c**, Comparison of two RIXS spectra recorded with the low (UD-NBCO, $\Delta E \sim 130$ meV) and high-energy resolution (UD-Bi2212, $\Delta E \sim 40$ meV) at $Q_{//} = 0.31$ r.l.u..

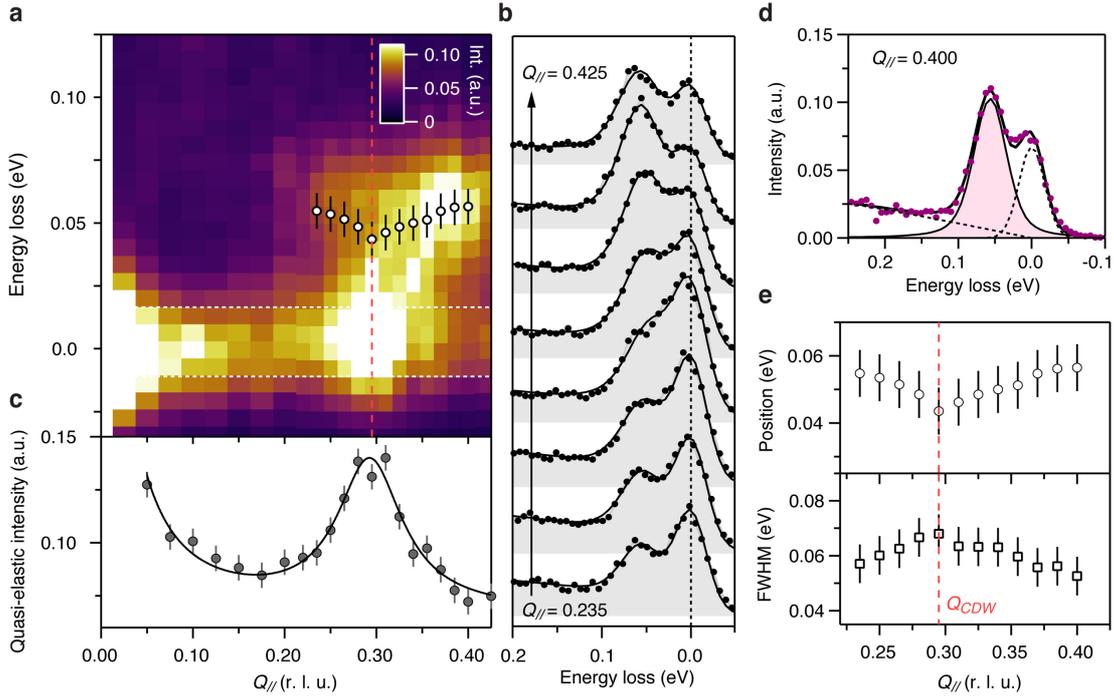

**Figure 2| CDW and phonons at 20K. a**, RIXS intensity map of the high energy resolution data ($\Delta E \sim 40$ meV) as a function of energy loss and $Q_{//}$. The white circles represent the fitted phonon dispersion. **b**, Energy loss spectra at selected momentum ranging from $Q_{//} = 0.235$ r.l.u. to $Q_{//} = 0.425$ r.l.u.. The fits (solid lines) are superimposed to the raw data (black circles). **c**, Averaged intensity in the quasi-elastic region, defined as the region between the two white dashed lines in **a**. The solid line is a Lorentzian fit to the data with a background consisting of a constant plus a Lorentzian to account for the tail of specular reflection peak at $Q_{//} = 0$. **d**, Demonstration of the quality of the fit of a RIXS spectrum at a representative momentum. The fitting process is described in the Method section. The phonon peak is highlighted in pink. **e**, Position and FWHM (full width at half maximum) of the measured phonon peak extracted from the fits. The red dashed lines on **a** and **e** indicate the $Q_{CDW}$. Error bars in **a** and **e** are estimated by the uncertainty in determining the zero energy loss. Those in **c** are determined by the noise level of the spectra.

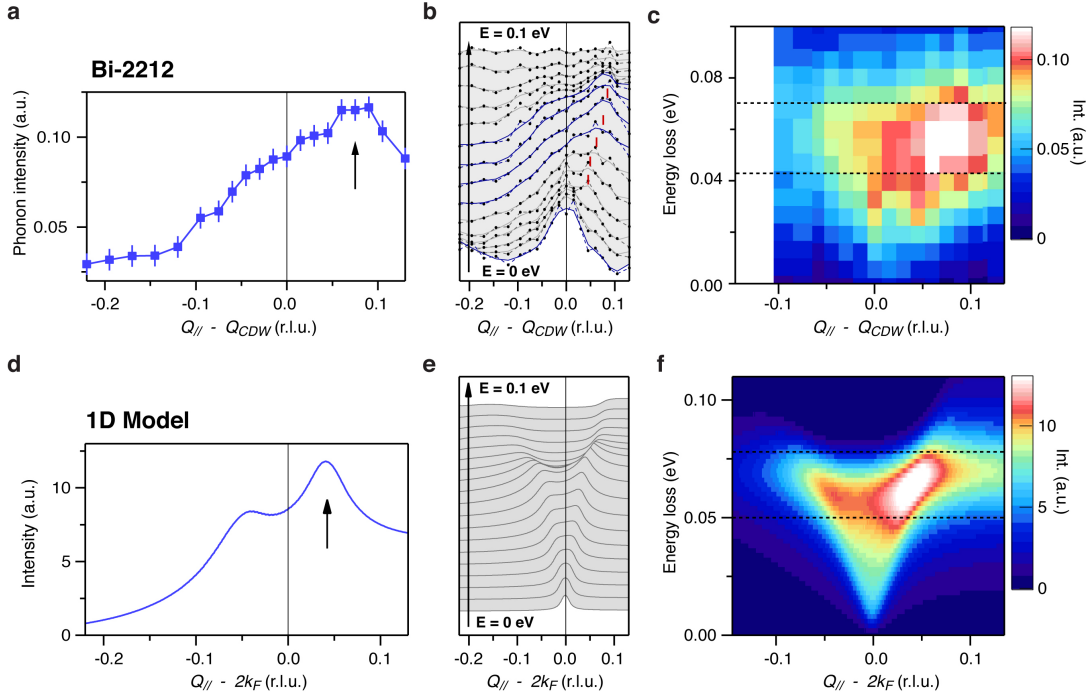

**Figure 3| CDW excitations in Bi2212 and calculated RIXS intensity in a 1D model. a**, RIXS phonon intensity extracted from the averaged intensity in the energy range between the two black dashed lines in **c**. The black arrow highlights the intensity anomaly. Error bars are estimated by the noise level of the spectra. **b,** Momentum distribution curves (MDCs) at fixed energy of the raw data (Fig. 2a) from $E = 0$ eV to $E = 0.1$ eV. Both the raw data (markers with dashed lines) and the smoothed curves (solid lines) are superimposed. The MDCs corresponding to the CDW and phonon energy regions are highlighted in blue. The red ticks serve as guides-to-the-eye for the connection between CDW and the phonon intensity anomaly. $Q_{CDW}$ represents the CDW wavevector. **c**, Elastic peak-subtracted intensity map. Calculations of RIXS phonon intensity for a 1D metallic system is shown in **d**, **e**, and **f**, with the same plotting format and notations used in panels **a–c**. $2k_F$ represents the perfect nesting wavevector of the 1D system. All the experimental (calculated) data are plotted as function of $Q_{//} - Q_{CDW}$ ($Q_{//} - 2k_F$).

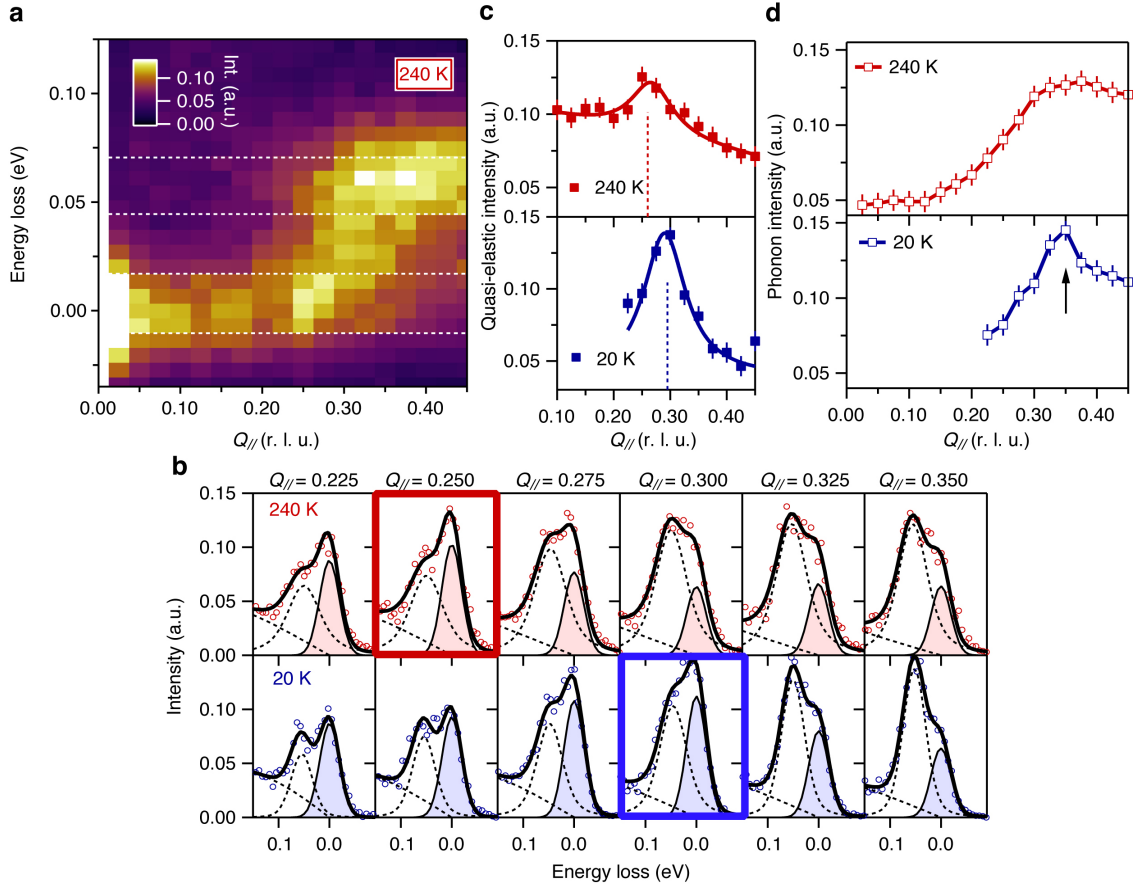

**Figure 4| Temperature dependence of the CDW and the phonon anomaly. a**, RIXS intensity map taken at $T = 240$ K. The white dashed lines define the quasi-elastic region (centered at zero energy loss) and the phonon energy region (centered at 60 meV) used for calculating the average intensity profile shown in **c** and **d**. **b**, Raw energy loss spectra (markers) and the corresponding fits (solid lines) of RIXS data taken at $T = 240$ K (top) and $T = 20$ K (bottom). The elastic peak fits are highlighted with filled areas. The RIXS spectra where the elastic peak intensity is maximal (*i.e.* the CDW position) are indicated by red and blue boxes for $T = 240$ K and $T = 20$ K, respectively. **c**, Averaged intensity of the quasi-elastic region (indicated in **a**) at 240 K (red) and 20 K (blue). The dashed lines indicate the position of the CDW. The RIXS intensity map taken at $T = 20$ K is presented in the Supplementary Fig. 3. **d**, Averaged intensity of the phonon energy regions, as defined in **a**, at 240 K and 20 K. The black arrow highlights the intensity anomaly at 20 K. Error bars in **c** and **d** are estimated by the noise level of the spectra.

Supplementary Information for

# Dispersive charge density wave excitations and temperature dependent commensuration in $Bi_2Sr_2CaCu_2O_{8+\delta}$

L. Chaix, G. Ghiringhelli, Y. Y. Peng, M. Hashimoto, B. Moritz, K. Kummer, N. B. Brookes, Y. He, S. Chen, S. Ishida, Y. Yoshida, H. Eisaki, M. Salluzzo, L. Braicovich, Z.-X. Shen[*], T. P. Devereaux[*], W.-S. Lee[*]

Correspondence to: leews@stanford.edu, zxshen@stanford.edu, tpd@stanford.edu

**This PDF file includes:**

Supplementary Text
Supplementary Figures
Supplementary References

## Supplementary Text

**Comparison with the CDW strength of YBCO**

RIXS spectra in the energy loss range of 1.5 - 3 eV contain orbital excitations within the Cu *3d* orbital manifold, the so-called *dd* excitations[34]. Since both Bi2212 and YBCO are double-layered systems, the orbital configurations of the Cu ions in the $CuO_2$ plane are similar. Thus, to the zero-th order, one could assume that the intensity of the *dd* excitation at $Q_{CDW}$ is similar between these two compounds. Thus, the intensity of *dd* can serve as an internal reference to compare the strength of CDW between these two systems. The peak height ratio of $I_{CDW}/I_{dd} \sim 0.6$ (*i.e.* the peak height of the CDW peak normalized to the one of the *dd* excitations) was extracted from the $YBa_2Cu_3O_{6.6}$ (YBCO) CDW peak profile ($T \sim 61$ K) presented in Ref. 4. In our data, the CDW peak profile is found to have a peak height ratio of $I_{CDW}/I_{dd} \sim 0.4$. Thus, the CDW peak height in Bi2212 is comparable with that in YBCO. It might be interesting to also compare the integrated CDW intensity, which presumably is more representative to the square of the amplitude of the CDW order parameter $\Delta^2$. To estimate the integrated peak intensity, we assume that the CDW is two-dimensional and isotropic (*i.e.* same correlation lengths along *a* and *b*). The CDW peak widths used for the estimation are 0.03 r.l.u. and 0.085 r.l.u. for YBCO and Bi2212, respectively. We found that $\Delta_{Bi2212}^2 / \Delta_{YBCO}^2 \sim 5$. We note that this estimate is crude, but nevertheless suggest that the CDW strength in Bi2212 is not weaker than that in $YBa_2Cu_3O_{6.6}$, despite of shorter correlation length.

**Comparison with the Cu-O bond stretching phonon of Bi2201 measured via IXS**

The Supplementary Fig. 4a presents a second derivative analysis of the raw data plotted in Fig. 2a (sample 1, $T = 20$ K). The data were first smoothed twice and then a second derivative was applied on each energy loss spectrum. From this fine analysis, we found that the energy-momentum dispersion of the 60 meV excitation seen in RIXS coincides with that of the Cu-O bond stretching phonon measured via non-resonant inelastic scattering (IXS) on another compound of the same family, the single-layer $Bi_2Sr_{1.6}La_{0.4}Cu_2O_{6+\delta}$ [23]. In addition, the momentum-dependence of the RIXS phonon intensity, which is weakest near the zone-center and gain intensity with increasing momentum, exhibiting an opposite trend with that measured via IXS (Supplementary Fig. 4b). This emphasizes an important aspect of the RIXS phonon cross-section – it directly reflects the momentum-dependent electron-phonon coupling strength[27] rather than the phonon self-energy that is probed by inelastic neutron scattering and IXS.

**Estimation of the quasi-particle-interference velocity**

We find that the quasi-particle-interference (QPI) velocities, $V_{QPI}$ ~ 0.2 - 0.4 eV•Å, which is deduced from linear fits to the dispersions of the QPI wavevectors $q_1$ from the Fourier transformed STM conductance maps reported in literatures[8,35,36]. In particular, we also estimate the QPI velocity from the energy-resolved STM conductance map of the Bi2212 sample having a doping concentration similar to our sample ($T_c$ = 45 K) plotted in Ref. 8. We found $V_{QPI}$ ~ 0.3 eV•Å at this doping concentration. In summary, $V_{QPI}$ is smaller than our observed $V_{CDW}$; thus they may not be directly related.

**Estimation of the Fermi velocity**

The Fermi velocities can be extracted from the analysis of the ARPES data. These measurements were performed at Stanford Synchrotron Radiation Lightsource (SSRL) beamline 5-4 with a SCIENTA R4000 analyzer and the measured UD50-Bi2212 (Bi2212 single-crystals with a $Tc$ = 50 K) samples were from the same sample growth batch as those used for the RIXS measurements. The data were taken using 18.4 eV photons with linear polarization perpendicular to the analyzer slit (measured cuts). The total energy and momentum resolution were set to ~10 meV and ~0.25°, respectively. The samples were cleaved and measured at 10 K with a pressure better than $3 \times 10^{-11}$ Torr. The Supplementary Fig. 5 presents a summary of the extraction of $V_F$. A constant energy intensity map integrated within $E_F \pm 5$ meV (*i.e.* the Fermi surface) is shown in Supplementary Fig. 5a and two dispersion (E–$k_y$) intensity maps where the separation between the two $k_F$'s corresponds to $Q \sim 0.28$ r.l.u. and $Q \sim 0.31$ r.l.u. (close to the CDW wavevector $Q_{CDW}$ of 0.3 r.l.u. measured in RIXS) are presented in the Supplementary Fig. 5b and c respectively. The momentum distribution curve (MDC) at each energy is fitted with a Lorentzian to obtain the electronic band dispersions (black lines on Supplementary Fig. 5b and c). To compare with the velocity extracted from the RIXS results, the $k_y$ component of $V_F$ (Supplementary Fig. 5d) is extracted from a linear fit of the MDC dispersions between 0 meV and -60 meV, as shown with red lines in Supplementary Fig. 5b and c. Thus, the $V_F$ at the Fermi surface where is connected by the $Q_{CDW}$ is ~ 2 eV•Å, which is significantly larger than our observed $V_{CDW}$.

# Supplementary Figures

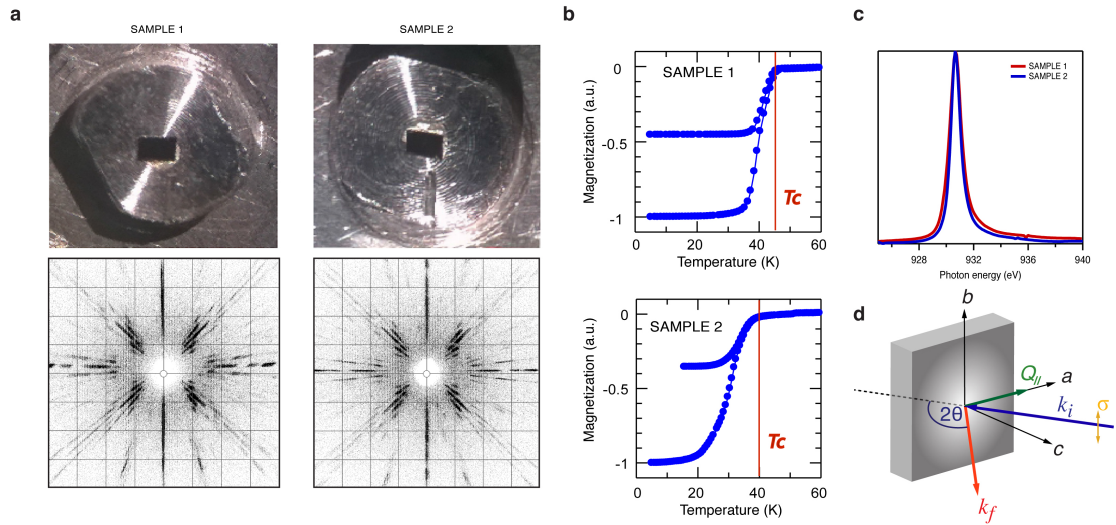

**Supplementary Figure 1| Sample Characterizations. a**, Top: photos of the samples (1 and 2) of the UD-Bi2212 used in our measurements. Bottom: Laue patterns for each sample. **b**, Magnetization measurements for the sample 1 and 2. The $T_C$ is determined by the onset of the superconducting transition. **c**, X-ray absorption spectra (XAS) near the Cu $L_3$-edge of the UD-Bi2212 sample 1 and sample 2. **d**, Scattering geometry used for this experiment. Linear vertical polarization (σ-polarization) of the incident beam was used. *a*, *b*, and *c* are the lattice axes of the sample. $k_i$ and $k_f$ represent the momentum of the incident and scattered photons, respectively. Since the electronic state in Bi2212 is quasi-two-directional, *i.e.* rather independent along the *c*-axis, the data shown in this report are plotted as a function of projected momentum transfer $Q_{//}$, *i.e.* projection of $Q = k_f - k_i$ on the CuO$_2$ plane. The reciprocal lattice units (r.l.u.) is denoted using pseudotetragonal Cu-O unit cell. The scattering angle 2θ is fixed at 149.5° for the RIXS measurements on Bi2212.

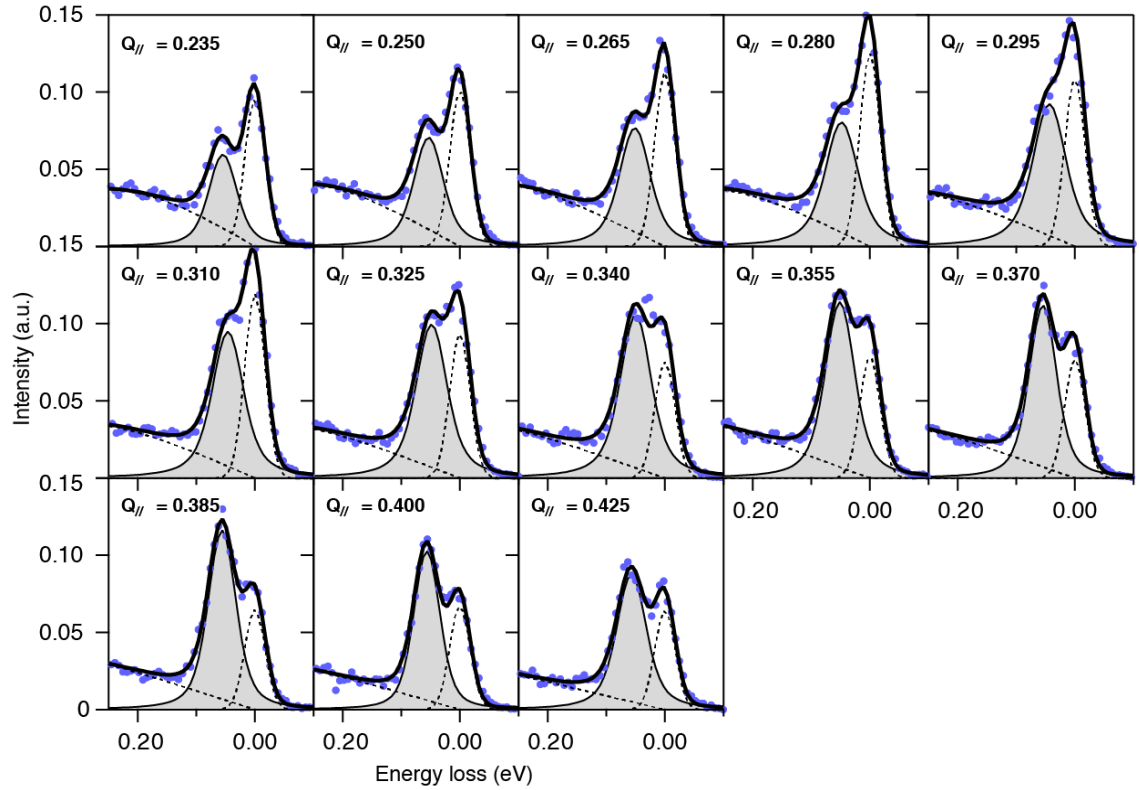

**Supplementary Figure 2|  Fits of all the RIXS spectra for the sample 1 at *T* = 20 K.** The model used takes into account a Gaussian for the elastic peak, a Lorentzian for the phonon and an anti-symmetrized Lorentzian[17] for the paramagnon, convoluted with the energy resolution (ΔE ~ 40 meV) via Gaussian convolution. The phonon peaks are highlighted in grey. The raw data are shown as markers.

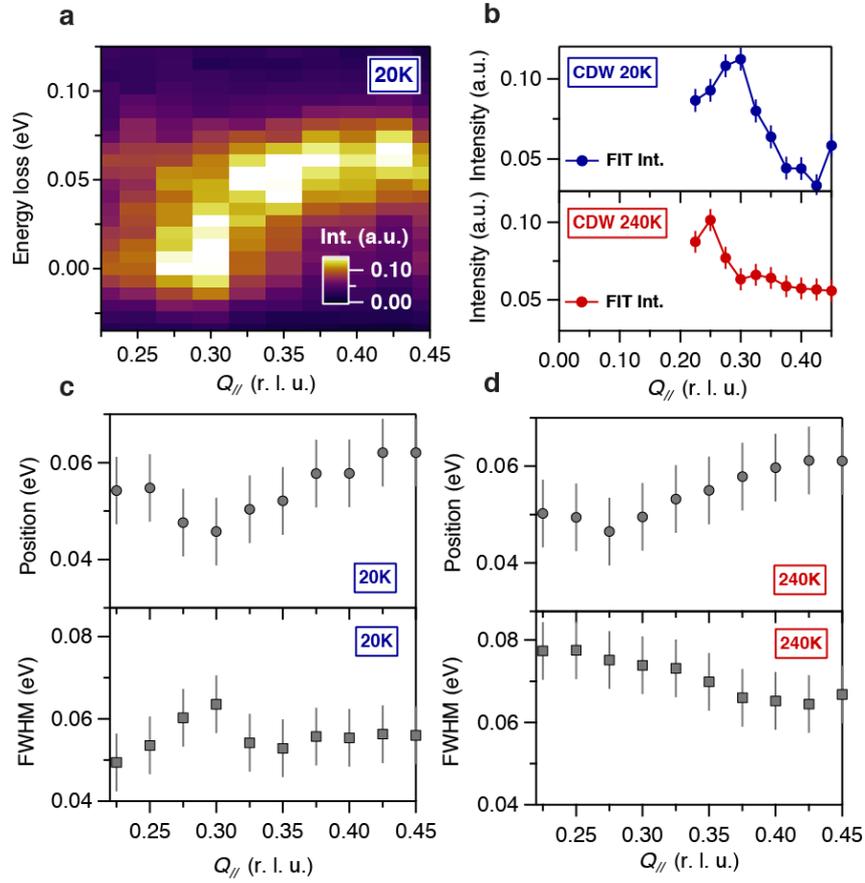

**Supplementary Figure 3| RIXS data on Sample 2. a**, RIXS intensity map collected at $T = 20$ K on sample 2, agreeing well with the measurement on sample 1 already shown in Fig. 2 of the main text. **b**, Intensity of the fitted elastic peak of the $T = 20$ K and $T = 240$ K data. They agree well with the average MDC at zero energy shown in the Fig. 4c of the main text. **c**, **d,** Position and measured FWHM (full width at half maximum) of the phonon peaks at $T = 20$ K and $T = 240$ K respectively. The parameters plotted on **b**, **c** and **d** are extracted from the fitting procedure presented on Fig. 4b. Error bars in **c** and **d** are estimated by the uncertainty of determining the zero energy loss. Those in **b** are determined by the noise level of the spectra.

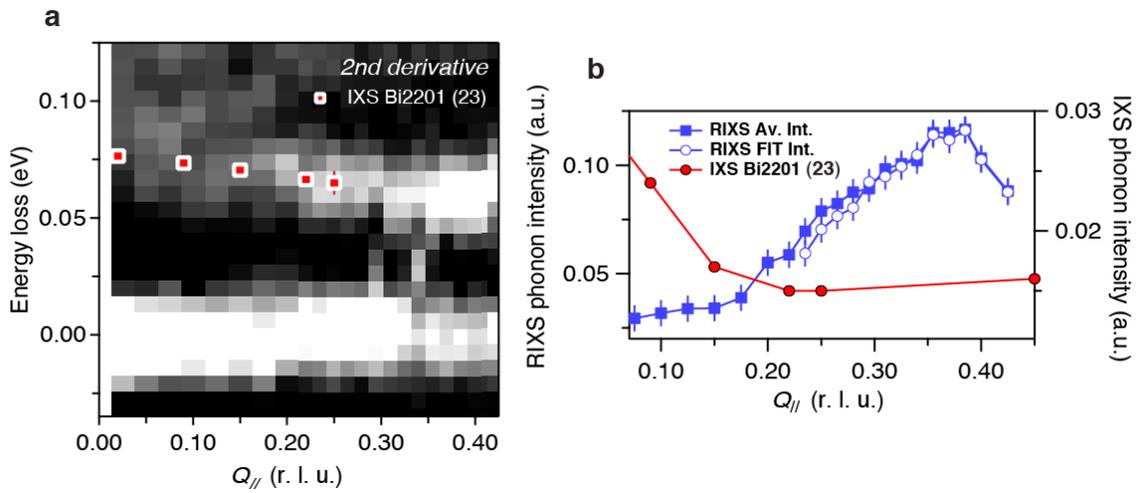

**Supplementary Figure 4| Comparison with IXS measurements. a,** Second derivative intensity map. To obtain this map, the data of Fig. 2 were first smoothed before applying the second derivative on each energy loss spectra. **b,** RIXS phonon intensities extracted from the fitted phonon intensity (open blue circles) and averaged intensity (blue squares) in the window indicated by the black dashed lines on Fig. 3c. Error bars are estimated by the noise level of the spectra. The red squares (red circles) on **a** (**b**) indicate the position (intensity) of the bond stretching phonon extracted from the IXS measurements on the single-layer $Bi_2Sr_{1.6}La_{0.4}Cu_2O_{6+\delta}$ (Bi2201) compound from Ref. 23.

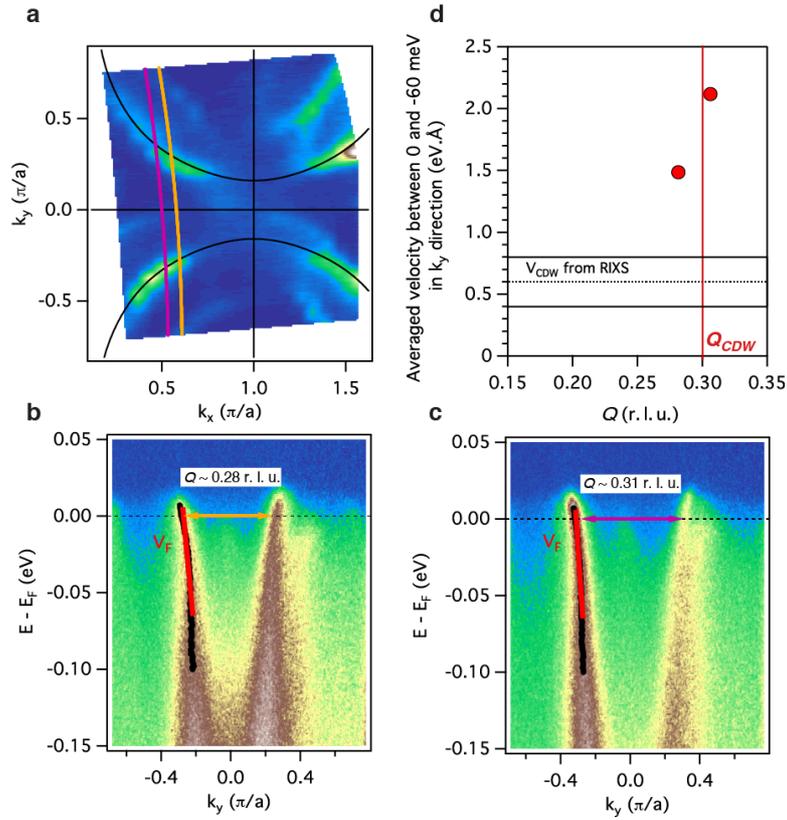

**Supplementary Figure 5| Fermi surface and band velocity measured by ARPES. a**, Fermi surface map of the UD50-Bi2212 sample ($T_C$ = 50 K). **b, c,** Electronic dispersion (E–$k_y$) intensity maps where the separation between the two $k_F$'s corresponds to $Q \sim 0.28$ r.l.u. (**b**) and $Q \sim 0.31$ r.l.u. (**c**), respectively (indicated by the solid lines on **a**). The black lines indicate the electronic band dispersions extracted from the Lorentzian fits of the momentum distribution curves. The red lines correspond to the linear fits of the MDC dispersions between 0 meV and -60 meV. **d**, Averaged Fermi velocity between 0 and -60 meV along $k_y$ as function of $Q$. The CDW wavevector found in our RIXS measurements is shown with the solid red line. The corresponding CDW excitation velocity of $0.6 \pm 0.2$ eV•Å is indicated with the dashed (averaged value) and solid (upper and lower limits) black lines.

**Supplementary Reference:**